\documentclass[12pt]{article}
\renewcommand{\theequation}{\thesection.\arabic{equation}}
\newcounter{subequation}[equation]
\makeatletter

\expandafter\let\expandafter\reset@font\csname reset@font\endcsname

\def\subeqnarray{\arraycolsep1pt
    \def\@eqnnum\stepcounter##1{\stepcounter{subequation}%
        {\reset@font\rm(\theequation\alph{subequation})}}
\jot5mm     \eqnarray}

\makeatother\newcommand{\newsection}[1]{\vspace{10mm}
\pagebreak[3]\addtocounter{section}{1}\setcounter{equation}{0}
\setcounter{subsection}{0}\setcounter{footnote}{0}

\begin{flushleft}{\Large\bf \thesection. #1}
\end{flushleft}\nopagebreak\medskip\nopagebreak}

\def\be{\begin{displaymath}}
\def\ee{\end{displaymath}}
\def\beq{\begin{equation}}
\def\eeq{\end{equation}}
\def\bea{\begin{eqnarray}}
\def\eea{\end{eqnarray}}
\def\l{\lambda}

\def\one#1{#1^{\raise5pt\hbox{$\scriptstyle\!\!\!\!1$}}\,{}}
\def\two#1{#1^{\raise5pt\hbox{$\scriptstyle\!\!\!\!2$}}\,{}}

\def\tilde{\widetilde}
\makeatletter
\def\binrel@#1{\begingroup
  \setboxz@h{\thinmuskip0mu
    \medmuskip\m@ne mu\thickmuskip\@ne mu
    \setbox\tw@\hbox{$#1\m@th$}\kern-\wd\tw@
    ${}#1{}\m@th$}%
  \edef\@tempa{\endgroup\let\noexpand\binrel@@
    \ifdim\wdz@<\z@ \mathbin
    \else\ifdim\wdz@>\z@ \mathrel
    \else \relax\fi\fi}%
  \@tempa
}
\let\binrel@@\relax
\def\overset#1#2{\binrel@{#2}%
  \binrel@@{\mathop{\kern\z@#2}\limits^{#1}}}
\def\underset#1#2{\binrel@{#2}%
  \binrel@@{\mathop{\kern\z@#2}\limits_{#1}}}
\makeatother
\newfont{\bbd}{msbm10 scaled\magstep1}

\newfont{\frak}{eufm10 scaled\magstep1}

\def\bt{B\"ack\-lund tran\-s\-for\-ma\-ti\-on}

\begin{document}
\begin{center}\LARGE\bf
B\"{a}cklund transformations for the sl(2) Gaudin magnet
\end{center}
\vskip1cm
\begin{center}
A N W  Hone,\dag\footnote[1]{E-mail: {\tt A.N.W.Hone@ukc.ac.uk}}
V B Kuznetsov\S\footnote[2]{E-mail: {\tt vadim@amsta.leeds.ac.uk}}
and O Ragnisco\ddag\footnote[3]{E-mail: {\tt ragnisco@fis.uniroma3.it}}
\vskip0.4cm
\dag Department of Mathematics, University of Kent, Canterbury CT2 7NF, UK\\
\S Department of Applied Mathematics,
University of Leeds, Leeds LS2 9JT, UK\\
\ddag Dipartimento di Fisica `E. Amaldi',
Universita degli Studi `Roma Tre', Roma, Italia
\end{center}
\vskip2cm
\begin{center}
\bf Abstract
\end{center}
Elementary, one- and two-point, B\"acklund transformations
are constructed for the generic case of the sl(2) Gaudin magnet.
The spectrality property is used to construct these explicitly
given, Poisson integrable maps which are time-discretizations
of the continuous flows with any Hamiltonian from the spectral
curve of the $2\times 2$ Lax matrix.
\vskip1cm

\newpage

\newsection{Introduction}
\setcounter{equation}{0}

\noindent
B\"{a}cklund transformations (BT's) are an essential tool
used to generate new solutions out of given solutions
to integrable equations. This is by now a well-developed
area, with elegant BT's having been found and studied for almost
all integrable hierarchies, see \cite{BTbks,AM}.

The theory of BT's for evolution equations had entered the subject
of finite-dimensional integrability through the discretization of
time variable(s). One of the most important and earliest accounts
on this subject is the papers by Veselov \cite{Ves} where
integrable Lagrange correspondences were introduced
as discrete-time analogs of
integrable continuous flows. Veselov clarified geometric
meaning of these correspondences
as finite shifts on Jacobians and gave several
important examples. Reader is referred to an extensive
literature which has appeared since then, see for instance
\cite{VSF,N,R,SS} and references therein.

In the present paper, following the approach of \cite{KS1}, we
look at BT's for finite-dimensional (Liouville) integrable systems
as special canonical transformations, thereby taking a Hamiltonian
point of view. We introduce and study several new properties
of BT's which appear to be very natural in such approach.

B\"{a}cklund transformations for finite-dimensional
integrable systems are defined in this paper as symplectic, or
more generally Poisson, integrable maps which are explicit
maps (rather than implicit multi-valued correspondences
of \cite{Ves}) and which can be viewed as
time discretizations of particular continuous flows.
The most characteristic features of such maps
are: i) BT's preserve the same set of integrals of motion as does
the continuous flow which they
discretize, ii) they depend on a (B\" acklund) parameter
$\lambda$ that specifies the corresponding shift on a Jacobian or on
a generalized Jacobian \cite{KV}, iii) a spectrality property holds
with respect to $\lambda$ and to the `conjugate' variable $\mu$,
which means that the point $(\lambda,\mu)$ belongs to the
spectral curve \cite{KS1,KS2}.

Because of the above properties, the constructed BT's are suitable as
explicit (symplectic) geometric integrators. Namely,
explicitness makes these maps to be pure iterative, while the
importance of the parameter $\lambda$ is that it allows
an adjustable discrete time step. The spectrality property
is strongly related to the simplecticness of the map.
Finally, numerical integrators which exactly preserve the
level set of integrals and at the same time are symplectic
proved to be impossible to find for generic Hamiltonian
dynamics \cite{Hairer}, but for integrable flows they
do exist and so are in demand.

In this paper we consider a generic (diagonal) case of the
sl(2) XXX Gaudin magnet which is an algebraic
completely integrable system.
We study the problem of constructing elementary
(one- and two-point)
B\"ack\-lund transformations for this system.
 In Section 3 we
construct an elementary (one-point) BT which gives an exact discretization of a
specific continuous flow. By making a two-point composite map
in Section 7 we are then able to discretize any of the independent
commuting flows with the Hamiltonians from the spectral curve
of the $2\times2$ Lax matrix.

\newsection{Gaudin magnet}
\setcounter{equation}{0}

\noindent
The sl(2) Gaudin magnet is derived from the Lax matrix
\beq
L(u)=\sum_{j=1}^n \frac1{u-a_j}\pmatrix{s_j^3&s_j^-\cr s_j^+&-s_j^3}
+\alpha\pmatrix{1&0\cr 0& -1}=\pmatrix{A(u)&B(u)\cr C(u) &-A(u)},
\label{lax}
\eeq
\beq
A(u)=\alpha+\sum_{j=1}^n \frac{s_j^3}{u-a_j}\,,\qquad
B(u)=\sum_{j=1}^n \frac{s_j^-}{u-a_j}\,,\qquad
C(u)=\sum_{j=1}^n \frac{s_j^+}{u-a_j}\,.
\eeq
Local variables in this model are generators of the direct sum of $n$
sl(2) spins, $s_j^3$, $s_j^\pm$, $j=1,\ldots,n$, with the following
Poisson brackets:
\beq
\{s_j^3,s_k^\pm\}=\mp \mbox{\rm i} \delta_{jk} s_k^\pm,\qquad
\{s_j^+,s_k^-\}=-2\mbox{\rm i} \delta_{jk} s_k^3.
\eeq
We denote Casimir operators (spins) as $s_j$:
\beq
s_j^2=\left(s_j^3\right)^2+s_j^+s_j^-.
\eeq
Fixing Casimirs $s_j$ we go to a symplectic leaf where the Poisson
bracket is non-degenerate, so that the symplectic manifold is a collection
of $n$ spheres.

Let us also introduce the total spin $\vec J$ which will be used later,
as follows:
\beq
J_3=\sum_{j=1}^n s_j^3,\qquad
J_+=\sum_{j=1}^n s_j^+,\qquad
J_-=\sum_{j=1}^n s_j^-.
\label{J}
\eeq

The Lax matrix (\ref{lax}) satisfies the linear $r$-matrix Poisson
algebra,
\beq
\{L_1(u),L_2(v)\}=[r(u-v),L_1(u)+L_2(v)],
\label{r}
\eeq
with the permutation matrix as the $r$-matrix
\beq
r(u-v)=\frac{\mbox{\rm i}}{u-v}\pmatrix{1&0&0&0\cr 0&0&1&0
\cr 0&1&0&0\cr 0&0&0&1}.
\label{rr}
\eeq
Here we use standard notations $L_1$ and $L_2$ for tensor products:
\beq
L_1(u)=L(u)\otimes \pmatrix{1&0\cr 0&1},\qquad
L_2(v)=\pmatrix{1&0\cr 0&1}\otimes L(v).
\eeq
Equation (\ref{r}) is equivalent to the following Poisson brackets for the
rational functions $A(u)$, $B(u)$, and $C(u)$:
\beq
\{A(u),A(v)\}=\{B(u),B(v)\}=\{C(u),C(v)\}=0,
\eeq
\beq
\{A(u),B(v)\}=\frac{\mbox{\rm i}}{u-v}\,(B(v)-B(u)),
\eeq
\beq
\{A(u),C(v)\}=\frac{-\mbox{\rm i}}{u-v}\,(C(v)-C(u)),
\eeq
\beq
\{C(u),B(v)\}=\frac{-2\mbox{\rm i}}{u-v}\,(A(v)-A(u)).
\eeq

The spectral curve $\Gamma$,
\beq
\Gamma: \quad \det(L(u)-v)=0,
\eeq
is a hyperelliptic, genus $n-1$ curve,
\beq
v^2=A^2(u)+B(u)C(u)=\alpha^2+\sum_{j=1}^n
\left(\frac{H_j}{u-a_j}+\frac{s_j^2}{(u-a_j)^2}\right),
\eeq
with the Hamiltonians (integrals of motion) $H_j$ of the form
\beq
H_j=\sum_{k\neq j} \frac{2s_j^3s_k^3+s_j^+s_k^-+s_j^-s_k^+}
{a_j-a_k}+2\alpha s_j^3.
\label{int}
\eeq
These are integrals of motion, or Hamiltonians, of the sl(2) Gaudin magnet,
which are Poisson commuting:
\beq
\{H_j,H_k\}=0,\qquad j,k=1,\ldots,n.
\eeq
Notice that there is one linear integral:
\beq
\sum_{j=1}^nH_j=2\alpha J_3.
\eeq

We can bring the curve $\Gamma$ into the canonical form by scaling
the variable $v\mapsto \hat v$:
\beq
\hat v=v\prod_{j=1}^n(u-a_j).
\eeq
The equation of the curve becomes
\be
\hat v^2=\left[\alpha^2+\sum_{j=1}^n\left(\frac{H_j}{u-a_j}+
\frac{s_j^2}{(u-a_j)^2}\right)\right] \prod_{j=1}^n (u-a_j)^2
\ee
\beq
=\alpha^2u^{2n}+f_1u^{2n-1}+f_2u^{2n-2}+\ldots+f_{2n}.
\label{cc}
\eeq
When $\alpha=0$ the genus of the curve drops to $n-2$,
because $f_1=0$ in such case. The Gaudin magnet then becomes
sl(2)-invariant: apart from integrals (\ref{int}) all three components
of the total spin $\vec J$ are integrals too,
\beq
\alpha=0:\qquad \{H_j,J_k\}=0,\qquad j=1,\ldots,n, \quad k=1,2,3.
\eeq
We will not consider this case,
but concentrate on the generic case of $\alpha\neq 0$ when there
is only one linear integral $f_1=2\alpha(J_3-\alpha\sum_{j=1}^na_j)$.
The latter case is called generic (diagonal) case of the sl(2) XXX Gaudin
magnet. It is known that all its flows are linearized on generalized
Jacobian of the hyperelliptic curve (\ref{cc}), see references in \cite{KV}.

\newsection{One-point basic map}
\setcounter{equation}{0}
The sl(2) Gaudin magnet with the $2\times 2$ Lax matrix (\ref{lax})
is within the class of systems that was considered recently in \cite{KV},
namely it belongs to the (even) case of the generalized Jacobian.
Hence, its B\"acklund transformations can be extracted from that paper.
However, we want to present here
an independent derivation of those BT's as well as to give more detailed
exposition of their various properties. The reader is referred
to \cite{KV} also for explanation of geometric meaning
of B\"acklund transformations.

A \bt\ should act on the Lax matrix as
a similarity transform:
\beq
L(u)\mapsto M(u) L(u) M(u)^{-1} \qquad \forall u,
\eeq
with some non-degenerate $2\times 2$ matrix $M(u)$, simply
because a BT should preserve the spectrum of $L(u)$.

Let us introduce new ($\,\tilde{}\;$-) notations for the up-dated variables
\beq
\tilde L(u)=\sum_{j=1}^n \frac1{u-a_j}\pmatrix{\tilde s_j{}^3
&\tilde s_j{}^-\cr \tilde s_j{}^+&-\tilde s_j{}^3}
+\alpha\pmatrix{1&0\cr 0& -1}=\pmatrix{\tilde A(u)&\tilde B(u)
\cr \tilde C(u) &-\tilde A(u)},
\label{lax2}
\eeq
\beq
\tilde A(u)=\alpha+\sum_{j=1}^n \frac{\tilde s_j{}^3}{u-a_j}\,,\qquad
\tilde B(u)=\sum_{j=1}^n \frac{\tilde s_j{}^-}{u-a_j}\,,\qquad
\tilde C(u)=\sum_{j=1}^n \frac{\tilde s_j{}^+}{u-a_j}\,,
\eeq
\beq
\{\tilde s_j{}^3,\tilde s_k{}^\pm\}=\mp \mbox{\rm i} \delta_{jk}
\tilde s_k{}^\pm,\qquad
\{\tilde s_j{}^+,\tilde s_k{}^-\}=-2\mbox{\rm i} \delta_{jk} \tilde s_k{}^3.
\eeq
We are looking for a Poisson map that intertwines two Lax matrices
$L(u)$ and $\tilde L(u)$:
\beq
M(u) L(u) = \tilde L(u) M(u)\qquad \forall u.
\label{dlax}
\eeq
Because spins $s_j$, $j=1,\ldots,n$, appear as coefficients
of the curve, they are not changed by the map, i.e. $\tilde s_j=s_j$.
Hence, we can talk about a symplectomorphism ($s_j=\mbox{const}$)
instead of a Poisson map.

Now we should choose an ansatz for the dependence of the matrix
$M(u)$ on the spectral parameter $u$. Let us fix the simplest
case of a {\it linear} function:
\beq
M(u)=M_1u+M_0.
\eeq
Taking limit $u\rightarrow\infty$ in (\ref{dlax}) we conclude
that $M_1$ must be diagonal. Moreover, the most elementary
(one-point) BT should correspond to the case when $\det M(u)$
has only one zero $u=\lambda$, which will lead
to having only one B\"acklund parameter
(cf. the spectrality property in \cite{KS1,KS2}).
So, we should choose either
\beq
M_1=\pmatrix{1&0\cr 0&0}\qquad \mbox{\rm or} \qquad
M_1=\pmatrix{0&0\cr 0&1}.
\eeq
We will consider the first case, as the second one will
produce a similar BT (cf. moving in discrete time in positive and
in negative direction). Finally, we arrive at the following
parameterisation of the unknown matrix $M(u)$:
\beq
M(u)=\pmatrix{u-\lambda+pq/\gamma & p \cr
                                  q&\gamma},
\label{M}
\eeq
with the $c$-number determinant
\beq
\det M(u)=\gamma(u-\lambda).
\eeq
Here the variables $p$ and $q$ are indeterminate dynamical
variables, but $\lambda$ and $\gamma$ are $c$-number
B\"acklund parameters\footnote[4]
{$M(u)$ comes from a $L$-operator of the quadratic
$r$-matrix algebra whose $\det$ is a Casimir, cf. \cite{S}.}.

Comparing asymptotics in $u$ in both sides of (\ref{dlax})
we readily get
\beq
p=\frac{J_-}{2\alpha}\,,\qquad q=\frac{\tilde J_+}{2\alpha}\,.
\label{p}
\eeq
If we want an explicit single-valued map from $L(u)$ to
$\tilde L(u)$
\beq
L(u)\mapsto \tilde L(u)=M(u) L(u) M^{-1}(u),
\label{m}
\eeq
then we must express $M(u)$, and therefore
$p$ and $q$, in terms of the old variables, i.e. the entries
of $L(u)$ only. There is however a problem, since from
(\ref{p}) we have only the expression for the $p$, but
the variable $q$ is given in terms of new, and therefore
{\it unknown} variable $\tilde J_+$. To overcome this difficulty
we will use an extra piece of data, namely the spectrality.
Apart from equations (\ref{dlax}) that our map satisfies, it will be
parameterised by the point $P$ on the curve $\Gamma$,
\beq
P=(\lambda,\mu)\in\Gamma.
\eeq
Notice that there are two points on the curve $\Gamma$,
$P=(\lambda,+\mu)$ and $Q=(\lambda,-\mu)$, corresponding
to a fixed $\lambda$ and sitting one above the other because
of the hyperelliptic involution:
\beq
(\lambda,\mu)\in\Gamma: \quad \det(L(\lambda)-\mu)=0
\quad \Leftrightarrow \quad \mu^2 +\det(L(\lambda))=0.
\eeq

Because $\det(M(\lambda))=0$ the matrix $M(\lambda)$ has
a one-dimensional kernel
\beq
M(\lambda)\Omega=\pmatrix{pq/\gamma&p\cr q&\gamma}\Omega=0,
\qquad \Omega=\pmatrix{\gamma\cr -q}.
\eeq
The equality
\beq
M(\lambda)L(\lambda)\Omega=\tilde L(\lambda) M(\lambda)\Omega=0
\eeq
implies that $L(\lambda)\Omega\sim\Omega$, so that $\Omega$ is
an eigenvector of $L(\lambda)$. Let us fix the corresponding point
of the spectrum as $P=(\lambda,\mu)$:
\beq
\pmatrix{A(\lambda)-\mu&B(\lambda)\cr
               C(\lambda)&-A(\lambda)-\mu}\pmatrix{\gamma\cr -q}=0.
\eeq
This gives us the formula for the variable $q$:
\beq
q=\gamma\,\frac{A(\lambda)-\mu}{B(\lambda)}=
-\gamma\,\frac{C(\lambda)}{A(\lambda)+\mu}\,.
\label{q}
\eeq
The two last expressions are equivalent since $(\lambda,\mu)\in\Gamma$.

Now, the formulas (\ref{m}), (\ref{M}), (\ref{p}), and (\ref{q}),
give a one-point Poisson integrable map ($\equiv$ one-point
BT) from $L(u)$ to $\tilde L(u)$.
The map is parameterised by one point $(\lambda,\mu)$
on the spectral curve $\Gamma$ (and by an extra parameter
$\gamma$).

Explicitly, it reads
\beq
\tilde A(u)=\frac{\gamma(u-\lambda+2pq/\gamma)A(u)
-q(u-\lambda+pq/\gamma)B(u)+p\gamma C(u)}{\gamma(u-\lambda)}\,,
\eeq
\beq
\tilde B(u)=\frac{(u-\lambda+pq/\gamma)^2B(u)
-2p(u-\lambda+pq/\gamma)A(u)-p^2 C(u)}{\gamma(u-\lambda)}\,,
\eeq
\beq
\tilde C(u)=\frac{\gamma^2C(u)+2q\gamma A(u)-q^2B(u)}{\gamma(u-\lambda)}\,.
\eeq
Equating residues at $u=a_j$ in both sides of the above equations,
we obtain the map in terms of the local spin variables:
\beq
\tilde s_j{}^3=\frac{\gamma(a_j-\lambda+2pq/\gamma)s_j^3
-q(a_j-\lambda+pq/\gamma)s_j^-+p\gamma s_j^+}
{\gamma(a_j-\lambda)}\,,
\label{loc1}
\eeq
\beq
\tilde  s_j{}^-=\frac{(a_j-\lambda+pq/\gamma)^2s_j^-
-2p(a_j-\lambda+pq/\gamma)s_j^3-p^2 s_j^+}
{\gamma(a_j-\lambda)}\,,
\label{loc2}
\eeq
\beq
\tilde s_j{}^+=\frac{\gamma^2s_j^++2q\gamma s_j^3-q^2s_j^-}
{\gamma(a_j-\lambda)}\,.
\label{loc3}
\eeq
Recall that $\alpha$, $a_j$ and $s_j$, $j=1,\ldots,n$, are parameters
of the model; $\gamma$ and $\lambda$ are parameters of the map;
and variables $p$ and $q$ are as follows:
\beq
p=\frac{J_-}{2\alpha}\,,\qquad
q=\gamma\,\frac{A(\lambda)-\mu}{B(\lambda)}=
-\gamma\,\frac{C(\lambda)}{A(\lambda)+\mu}\,,
\eeq
\beq
\mu^2=\alpha^2+\sum_{j=1}^n
\left(\frac{H_j}{\lambda-a_j}+\frac{s_j^2}
{(\lambda-a_j)^2}\right),
\eeq
\beq
H_j=\sum_{k\neq j} \frac{2s_j^3s_k^3+s_j^+s_k^-+s_j^-s_k^+}
{a_j-a_k}+2\alpha s_j^3.
\eeq

\newsection{BT as a discrete-time map}
\setcounter{equation}{0}
In this Section we will show that the BT constructed above
can be seen as a time-discreti\-za\-tion of a specific Hamiltonian
flow where the parameter $\lambda$ plays a role
of inversion of the time step.

Consider the limit $\lambda\rightarrow\infty$ then
\beq
\mu=\alpha+O\left(\frac1\lambda\right).
\eeq
Assume that
\beq
\gamma=-\lambda+\gamma_0+O\left(\frac1\lambda\right),
\eeq
then we have the following expansion for the matrix $M(u)$:
\beq
M(u)=-\lambda\left(1-\frac1{2\lambda}M_0(u)\right)
+O\left(\frac1\lambda\right).
\eeq
The equation of the map, $M(u)L(u)=\tilde L(u) M(u)$,
turns in this limit into the Lax pair of a continuous flow:
\beq
\dot L(u)=[L(u),M_0(u)],\qquad M_0(u)=\pmatrix{u-\gamma_0&
J_-/\alpha\cr J_+/\alpha&-u+\gamma_0},
\label{flow}
\eeq
where $1/(2\lambda)$ is a time-step and $\dot L(u)\equiv
\lim_{\lambda\rightarrow\infty}2\lambda(\tilde L(u)-L(u))$
is the time-derivative.

The flow (\ref{flow}) is a Hamiltonian flow
\beq
\dot L(u)=\{H,L(u)\},
\eeq
with the Hamiltonian function $H$ as
\beq
H=\frac{\mbox{\rm i}}{\alpha}\left(J_+J_-
+2\alpha\sum_{j=1}^n(a_j-\gamma_0)s_j^3\right).
\eeq
Therefore, the constructed \bt\ is a two-parameter
($\lambda,\gamma$) time-discreti\-za\-tion
of this continuous flow.

\newsection{Symplecticity}
\setcounter{equation}{0}

\noindent
In this Section we give a simple proof of symplecticity
of the constructed map by finding an explicit generating
function of the corresponding canonical transformation
from the old to new variables.

First, because the spin variables (Casimirs) do not change,
\beq
s_j^2=\left(s_j^3\right)^2+s_j^+s_j^-=
\left(\tilde s_j{}^3\right)^2+\tilde s_j{}^+\tilde s_j{}^-,
\eeq
we can exclude the variables $s_j^+$ and $\tilde s_j{}^-$,
$j=1,\ldots,n$,
\beq
s_j^+=\frac{s_j^2-\left(s_j^3\right)^2}{s_j^-}\,,\qquad
\tilde s_j{}^-=\frac{s_j^2-\left(\tilde s_j{}^3\right)^2}
{\tilde s_j{}^+}\,,
\eeq
expressing everything in terms of $2n$ `canonical'
variables $(s_j^3,s_j^-)_{j=1}^n$ and
$(\tilde s_j{}^3,\tilde s_j{}^+)_{j=1}^n$ with
the following Poisson brackets:
\beq
\{s_j^3,s_k^-\}=\mbox{\rm i}\delta_{jk}s_k^-,\qquad
\{\tilde s_j{}^3,\tilde s_k{}^+\}=-\mbox{\rm i}\delta_{jk}
\tilde s_k{}^+.
\eeq
We want to represent our \bt\ as a canonical transformation
defined by the generating function $F(\tilde s{\,}^+|s^-)\equiv
F_{\lambda,\gamma}(\tilde s{\,}^+|s^-)$ such that
\beq
s_j^3=\mbox{\rm i} s_j^-\,\frac{\partial F(\tilde s{\,}^+|s^-)}
{\partial s_j^-}\,, \qquad
\tilde s_j{}^3=\mbox{\rm i} \tilde s_j{}^+\,
\frac{\partial F(\tilde s{\,}^+|s^-)}{\partial \tilde s_j{}^+}\,.
\label{spe}
\eeq
Notice that we have chosen the arguments of the generating
function as $(\tilde s_j{}^+|s_j^-)_{j=1}^n$. Because the
symplecticity property does not depend on the choice
of the arguments of its generating function,
these arguments are fixed in order to get a simpler expression
for the function $F_{\lambda,\gamma}$ (recall that the variables
$p$ and $q$ (\ref{p}) depend exactly on these variables).

Rewrite now the equations of the map (\ref{loc1})--(\ref{loc3}) in the form
\beq
\left(\gamma s_j^3-\frac{\tilde J_+}{2\alpha}\,s_j^-\right)^2
+\gamma(a_j-\l)\tilde s_j{}^+s_j^--\gamma^2s_j^2=0,
\eeq
\beq
\left(\gamma \tilde s_j{}^3-\frac{J_-}{2\alpha}\,\tilde s_j{}^+\right)^2
+\gamma(a_j-\l)\tilde s_j{}^+s_j^--\gamma^2s_j^2=0.
\eeq
Resolving them with respect to $s_j^3$ and $\tilde s_j{}^3$,
we obtain
\beq
s_j^3=\frac{\tilde J_+}{2\alpha\gamma}\,s_j^-+z_j,\qquad
\tilde s_j{}^3=\frac{J_-}{2\alpha\gamma}\,\tilde s_j{}^++z_j,
\label{two}
\eeq
\beq
z_j^2=s_j^2-\frac{a_j-\l}\gamma\, \tilde s_j{}^+s_j^-,\qquad
j=1,\ldots,n.
\eeq
It is now easy to check that the function
\beq
F_{\lambda,\gamma}(\tilde s{\,}^+|s^-)=-\mbox{\rm i}\,\frac{\tilde J_+J_-}
{2\alpha\gamma}
-\mbox{\rm i} \sum_{j=1}^n\left(2z_j+s_j\log\frac{z_j-s_j}{z_j+s_j}\right)
\label{theF}
\eeq
satisfies equations (\ref{spe}). Thereby symplecticity of the map is proven.

\newsection{Spectrality}
\setcounter{equation}{0}

\noindent
The map depends on two parameters, $\lambda$ and $\gamma$.
Let us first concentrate on its $\lambda$-dependence.

Spectrality, which was introduced in \cite{KS1}, is an interesting
property of B\"acklund transformations. It usually holds for
any \bt\ which has a parameter. Technically, it means that
the components of the point $P=(\lambda,\mu)\in\Gamma$
which parameterises the map are conjugated variables, namely:
\beq
\mu=\frac{\partial F_{\lambda,\gamma}(\tilde s{\,}^+|s^-)}{\partial \lambda}\,.
\label{prop}
\eeq
To prove this formula, use (\ref{p}) and (\ref{q}) to find the formula
for the $\mu$,
\beq
\mu=A(\l)-\frac{\tilde J_+}{2\alpha\gamma}\,B(\l).
\eeq
Now, with the help of (\ref{two}) and (\ref{theF}) we
easily check the needed formula for the spectrality
property (\ref{prop}).

A new, comparing to \cite{KS1}, observation is that there
is also an analogous property with respect to the parameter
$\gamma$, only now it is the integral $J_3$ that plays
the role of the conjugated variable:
\beq
J_3=-\gamma \,\frac{\partial F_{\lambda,\gamma}(\tilde s{\,}^+|s^-)}{\partial
\gamma}\,.
\label{second}
\eeq
The proof is very simple, once we notice that (\ref{two}) entails
\beq
J_3=\frac{\tilde J_+J_-}{2\alpha\gamma}+\sum_{j=1}^nz_j.
\eeq
Concluding this Section we want to remark that
because of the second `spectrality' property (\ref{second}),
which was somehow built into the \bt\ from the very beginning,
one could recover the generating function of the
map just taking one integral,
\beq
F_{\lambda,\gamma}(\tilde s{\,}^+|s^-)=\int^\gamma
\left(-\frac{J_3}\gamma \right) \mbox{\rm d} \gamma + \mbox{\rm const},
\eeq
without needing to solve the system of $2n$ differential equations
(\ref{two}).

\newsection{Inverse map and a two-point map}
\setcounter{equation}{0}

\noindent
In this Section we will first construct the inverse
\bt\ and then use it to derive the two-point
B\"acklund transformation which will be a
composition of the direct map parameterised
by the point $P_1=(\lambda_1,\mu_1)\in\Gamma$
and the inverse map parameterised by the
point $Q_2=(\lambda_2,-\mu_2)\in\Gamma$.

\subsection{The inversion of the \bt\ }

\noindent
Let us call the direct map by $B_P$.
The inverse map acts from $\tilde L(u)$ to $L(u)$. We can rewrite
the equations for the $B_P$ in the inverse form
\beq
M^\wedge(u)\tilde L(u)=L(u)M^\wedge(u), \qquad
M^\wedge(u)=\pmatrix{\gamma&-p\cr -q&u-\l+pq/\gamma}.
\eeq
To define the inverse map we must find expressions
for the co-factor matrix $M^\wedge(u)$, or for the variables
$p$ and $q$, in terms of $\,\tilde{\;\;}$-variables, i.e. in terms
of the entries of $\tilde L(u)$.
We have already the expressions (\ref{p}),
\beq
p=\frac{J_-}{2\alpha}\,,\qquad q=\frac{\tilde J_+}{2\alpha}\,,
\label{pp}
\eeq
which define $q$. To obtain the formula for the variable
$p$ we will use again the spectrality property.
The matrix $M^\wedge(\lambda)$ has a one-dimensional kernel
$\tilde\Omega$,
\beq
M^\wedge(\lambda)\tilde\Omega=
\pmatrix{\gamma&-p\cr -q&pq/\gamma}\tilde \Omega=0,
\qquad \tilde \Omega=\pmatrix{p\cr \gamma}.
\eeq
The main difference comparing to the formulas of the
direct map is that
the inverse map will be para\-metrized by the point
$Q=(\lambda,-\mu)\in \Gamma$, not the $P=(\lambda,\mu)\in\Gamma$.
Therefore, $\tilde
\Omega$ is an eigenvector of the matrix $\tilde L(u)$
with the eigenvalue $Q=(\lambda,-\mu)$:
\beq
\pmatrix{\tilde A(\lambda)+\mu&\tilde B(\lambda)\cr
               \tilde C(\lambda)&-\tilde A(\lambda)+\mu}\pmatrix{p\cr \gamma}=0.
\eeq
This gives us the needed formula for the variable $p$,
\beq
p=\gamma\,\frac{\tilde A(\lambda)-\mu}{\tilde C(\lambda)}=
-\gamma\,\frac{\tilde B(\lambda)}{\tilde A(\lambda)+\mu}\,.
\label{qq}
\eeq
To prove that this \underline{does} indeed give the inverse map,
we have to show that the two formulas, (\ref{q})
and (\ref{qq}), define in fact the same variable $\mu$:
\beq
\mu=\tilde A(\l)-\frac{p}\gamma\;\tilde C(\l)
\stackrel{?}{=}A(\l)-\frac{q}\gamma\;
B(\l).
\eeq
It is easy to see that this equation is the $(11)$-element
of the matrix identity:
\beq
M^\wedge(\l)\tilde L(\l)=L(\l)M^\wedge(\l).
\eeq
We will denote as ${\cal B}_Q$ the map which is inverse
to the map $B_P$.
Generally speaking, we have constructed 4 different maps,
$B_P$, $B_Q$,  ${\cal B}_Q$, and ${\cal B}_P$,
with two pairs of maps which are inverse to each other:
\beq
{\cal B}_Q \circ B_P=B_P \circ {\cal B}_Q=
{\cal B}_P \circ B_Q=B_Q \circ {\cal B}_P=\mbox{\rm Id}.
\eeq

\subsection{The two-point map $B_{P_1,Q_2}$}

We now construct a composite map which is a product
of the map $B_{P_1}\equiv B_{(\l_1,\mu_1)}$ and
${\cal B}_{Q_2}\equiv {\cal B}_{(\l_2,-\mu_2)}$:
\beq
B_{P_1,Q_2}={\cal B}_{Q_2} \circ B_{P_1}.
\eeq
The second parameter of the basic map, namely the $\gamma$,
is taken the same in both maps, so $\gamma_1=\gamma_2$.
Obviously, when $\l_1=\l_2$ (and $\mu_1=\mu_2$) this composite
map will turn into an identity map.

The first map $B_{P_1}$ reads as follows:
\beq
M_1(u)L(u)=\tilde L(u)M_1(u), \qquad
M_1(u)=\pmatrix{u-\l_1+p_1q_1/\gamma&p_1\cr q_1&\gamma},
\eeq
where the formulas for the variables $p_1$ and $q_1$ are
\beq
p_1=\frac{J_-}{2\alpha}=
\gamma\,\frac{\tilde A(\lambda_1)-\mu_1}{\tilde C(\lambda_1)}=
-\gamma\,\frac{\tilde B(\lambda_1)}{\tilde A(\lambda_1)+\mu_1}\,,
\label{p1}
\eeq
\beq
q_1=\frac{\tilde J_+}{2\alpha}=
\gamma\,\frac{A(\lambda_1)-\mu_1}{B(\lambda_1)}=
-\gamma\,\frac{C(\lambda_1)}{A(\lambda_1)+\mu_1}\,.
\label{q1}
\eeq
The second map $B_{Q_2}$ reads as follows:
\beq
M_2(u)\tilde L(u)=\tilde{\tilde L}(u)M_2(u), \qquad
M_2(u)=\pmatrix{\gamma &-p_2\cr -q_2&u-\lambda_2+p_2q_2/\gamma},
\eeq
where the formulas for the variables $p_2$ and $q_2$ are
\beq
p_2=\frac{\tilde{\tilde J}_-}{2\alpha}=
\gamma\,\frac{\tilde A(\lambda_2)-\mu_2}{\tilde C(\lambda_2)}=
-\gamma\,\frac{\tilde B(\lambda_2)}{\tilde A(\lambda_2)+\mu_2}\,,
\label{p2}
\eeq
\beq
q_2=\frac{\tilde J_+}{2\alpha}=
\gamma\,\frac{\tilde{\tilde A}(\lambda_2)-\mu_2}{\tilde{\tilde B}(\lambda_2)}=
-\gamma\,\frac{\tilde{\tilde C}(\lambda_2)}{\tilde{\tilde A}(\lambda_2)+\mu_2}\,.
\label{q2}
\eeq
Notice that $q_1$ is equal to $q_2$, hence we omit the sub-index,
$q_1=q_2=q$.

The composite map $B_{P_1,Q_2}$ acts from $L(u)$ to $\tilde{\tilde L}(u)$,
\beq
M(u)L(u)=\tilde{\tilde L}(u)M(u),
\eeq
\beq
M(u)=\frac1\gamma\; M_2(u)M_1(u)=
\pmatrix{u-\l_1+\frac{q}\gamma\,(p_1-p_2)& p_1-p_2\cr
                \frac{q}\gamma\,(\l_1-\l_2-\frac{q}\gamma\,(p_1-p_2))&
        u-\l_2-\frac{q}\gamma\,(p_1-p_2)}.
\label{last1}
\eeq

In order to get rid of the intermediate $\tilde{\;\;}$-variables, we will
use the spectrality property with respect to two points, $P_1=(\l_1,\mu_1)$
and $Q_2=(\l_2,-\mu_2)$. Obviously, both spectralities are still valid
after composing the maps. For the point $P_1$ we get the following
equations:
\be
M(\l_1)\Omega_1=0,\qquad \Omega_1=\pmatrix{\gamma\cr -q},
\qquad L(\l_1)\Omega_1=\mu_1\Omega_1\quad\Rightarrow
\ee
\beq
q=\gamma\,\frac{A(\lambda_1)-\mu_1}{B(\lambda_1)}=
-\gamma\,\frac{C(\lambda_1)}{A(\lambda_1)+\mu_1}\,;
\label{x11}
\eeq
\be
M^\wedge(\l_1){\tilde{\tilde\Omega}}_1=0,\qquad
{\tilde{\tilde\Omega}}_1=\pmatrix{p_1-p_2\cr \l_1-\l_2-\frac{q}
{\gamma}\,(p_1-p_2)},\qquad
{\tilde{\tilde L}}(\l_1){\tilde{\tilde\Omega}}_1=
-\mu_1{\tilde{\tilde\Omega}}_1\quad\Rightarrow
\ee
\beq
p_1-p_2=\frac{\gamma(\l_2-\l_1){\tilde{\tilde B}}(\l_1)}
{\gamma\left({\tilde{\tilde A}}(\l_1)+\mu_1\right)-q{\tilde{\tilde B}}(\l_1)}
=\frac{\gamma(\l_1-\l_2)\left({\tilde{\tilde A}}(\l_1)-\mu_1\right)}
{q\left({\tilde{\tilde A}}(\l_1)-\mu_1\right)+\gamma{\tilde{\tilde C}}(\l_1)}\,.
\label{x22}
\eeq
For the point $Q_2$ we get the second set of equations:
\be
M(\l_2)\Omega_2=0,\qquad \Omega_2=
\pmatrix{p_1-p_2\cr \l_1-\l_2-\frac{q}{\gamma}\,(p_1-p_2)},
\qquad L(\l_2)\Omega_2=-\mu_2\Omega_2\quad\Rightarrow
\ee
\beq
p_1-p_2=\frac{\gamma(\l_2-\l_1)B(\l_2)}
{\gamma\left(A(\l_2)+\mu_2\right)-qB(\l_2)}
=\frac{\gamma(\l_1-\l_2)\left(A(\l_2)-\mu_2\right)}
{q\left(A(\l_2)-\mu_2\right)+\gamma C(\l_2)}\,;
\label{x33}
\eeq
\be
M^\wedge(\l_2){\tilde{\tilde\Omega}}_2=0,\qquad
{\tilde{\tilde\Omega}}_2=\pmatrix{\gamma\cr -q},\qquad
{\tilde{\tilde L}}(\l_2){\tilde{\tilde\Omega}}_2=
\mu_2{\tilde{\tilde\Omega}}_2\quad\Rightarrow
\ee
\beq
q=\gamma\,\frac{\tilde{\tilde A}(\lambda_2)-\mu_2}{\tilde{\tilde B}(\lambda_2)}=
-\gamma\,\frac{\tilde{\tilde C}(\lambda_2)}{\tilde{\tilde A}(\lambda_2)+\mu_2}\,.
\label{x44}
\eeq

Equations (\ref{x11}) and (\ref{x44}) are already known to us
(cf. (\ref{q1}) and (\ref{q2})). The formulas (\ref{x22}) and
(\ref{x33}) for the variable $p_1-p_2$ are new. They are
equivalent to the formulas
(\ref{p1}) and (\ref{p2}) expressed in terms of entries of $L(u)$ and
$\tilde{\tilde L}(u)$.

Concluding, we have constructed a two-point \bt\ which
is factorised to two one-point B\"acklund transformations
and which is  explicitly given, together with its inverse, by the
formulas:
\beq
M(u)L(u)=\tilde{\tilde L}(u)M(u),\qquad
M(u)=\pmatrix{u-\l_1+xX&X\cr -x^2X+(\l_1-\l_2)x&u-\l_2-xX},
\label{matrixM}
\eeq
where
\beq
x:=\frac{A(\lambda_1)-\mu_1}{B(\lambda_1)}=
-\frac{C(\lambda_1)}{A(\lambda_1)+\mu_1}
=\frac{\tilde{\tilde A}(\lambda_2)-\mu_2}{\tilde{\tilde B}(\lambda_2)}=
-\frac{\tilde{\tilde C}(\lambda_2)}{\tilde{\tilde A}(\lambda_2)+\mu_2}\,,
\eeq
\beq
X:=\frac{(\l_2-\l_1)B(\l_1)B(\l_2)}{B(\l_1)(A(\l_2)+\mu_2)-B(\l_2)(A(\l_1)-\mu_1)}\;\
;
\eeq
\beq
=\frac{(\l_1-\l_2)B(\l_1)(A(\l_2)-\mu_2)}{(A(\l_1)-\mu_1)(A(\l_2)-\mu_2)
+B(\l_1)C(\l_2)}
\eeq
\beq
=\frac{(\l_2-\l_1)B(\l_2)(A(\l_1)+\mu_1)}{(A(\l_1)+\mu_1)(A(\l_2)+\mu_2)
+B(\l_2)C(\l_1)}
\eeq
\beq
=\frac{(\l_1-\l_2)(A(\l_1)+\mu_1)(A(\l_2)-\mu_2)}{(A(\l_1)+\mu_1)C(\l_2)
-(A(\l_2)-\mu_2)C(\l_1)}
\eeq
\beq
=\frac{(\l_2-\l_1)\tilde{\tilde B}(\l_2)\tilde{\tilde B}(\l_1)}
{\tilde{\tilde B}(\l_2)\left(\tilde{\tilde A}(\l_1)+\mu_1\right)-
\tilde{\tilde B}(\l_1)\left(\tilde{\tilde A}(\l_2)-\mu_2\right)}
\eeq
\beq
=\frac{(\l_1-\l_2)\tilde{\tilde B}(\l_2)\left(\tilde{\tilde A}(\l_1)-\mu_1\right)}
{\left(\tilde{\tilde A}(\l_2)-\mu_2\right)\left(\tilde{\tilde A}(\l_1)-\mu_1\right)
+\tilde{\tilde B}(\l_2)\tilde{\tilde C}(\l_1)}
\eeq
\beq
=\frac{(\l_2-\l_1)\tilde{\tilde B}(\l_1)\left(\tilde{\tilde A}(\l_2)+\mu_2\right)}
{\left(\tilde{\tilde A}(\l_2)+\mu_2\right)\left(\tilde{\tilde A}(\l_1)+\mu_1\right)
+\tilde{\tilde B}(\l_1)\tilde{\tilde C}(\l_2)}
\eeq
\beq
=\frac{(\l_1-\l_2)\left(\tilde{\tilde A}(\l_2)+\mu_2\right)
\left(\tilde{\tilde A}(\l_1)-\mu_1\right)}{\left(\tilde{\tilde A}(\l_2)+\mu_2\right)
\tilde{\tilde C}(\l_1)
-\left(\tilde{\tilde A}(\l_1)-\mu_1\right)\tilde{\tilde C}(\l_2)}\,.
\eeq
The above formulas give several equivalent expressions for the variables
$x$ and $X$ since the points $(\l_1,\mu_1)$ and $(\l_2,-\mu_2)$
belong to the spectral curve $\Gamma$, i.e. are bound by the following relations:
\beq
\mu_1^2=A^2(\l_1)+B(\l_1)C(\l_1),\qquad \mu_2^2=A^2(\l_2)
+B(\l_2)C(\l_2),
\eeq
\beq
\mu_1^2={\tilde{\tilde A}\,}^2(\l_1)+\tilde{\tilde B}(\l_1)\tilde{\tilde C}(\l_1),
\qquad \mu_2^2={\tilde{\tilde A}\,}^2(\l_2)+\tilde{\tilde B}(\l_2)\tilde{\tilde
C}(\l_2).
\eeq

\subsection{Two-point map as a discrete-time map}

We will see in this Section that the two-point map constructed
above is a one-parameter, $\l_1$, time-discretization
of a family of flows parameterised by the point
$Q_2=(\l_2,-\mu_2)$, with the difference $\l_1-\l_2$
playing the role of the time-step.

Indeed, consider the limit $\l_1\rightarrow \l_2$,
\beq
\l_1=\l_2+\varepsilon,\qquad \varepsilon\rightarrow0.
\eeq
It is easy to see from the formulas of the previous subsection that
\beq
x=x_0+O(\varepsilon),\qquad
x_0=\frac{A(\lambda_2)-\mu_2}{B(\lambda_2)}=
-\frac{C(\lambda_2)}{A(\lambda_2)+\mu_2}
\eeq
and
\beq
X=\varepsilon X_0+O(\varepsilon^2),\qquad
X_0=-\frac{B(\l_2)}{2\mu_2}\,.
\eeq
Then we derive that the matrix $M(u)$ has the following
asymptotics:
\beq
M(u)=(u-\l_2)\left(1-\frac{\varepsilon}{2\mu_2(u-\l_2)}\,
\pmatrix{A(\l_2)+\mu_2&B(\l_2)\cr C(\l_2)&-A(\l_2)+\mu_2}
\right)+O(\varepsilon^2).
\eeq
If we now define the time-derivative $\dot L(u)$ as
\beq
\dot L(u)=\lim_{\varepsilon\rightarrow0}\frac{\tilde{\tilde L}(u)-L(u)}\varepsilon,
\eeq
then in the limit we obtain from the equation of the map,
$M(u)L(u)=\tilde{\tilde L}(u)M(u)$,
the Lax equation for a corresponding continuous
flow that our \bt\ discretizes, namely:
\beq
\dot L(u)=\left[L(u),\frac{L(\l_2)}{2\mu_2(u-\l_2)}\right].
\label{rlax}
\eeq
This is a Hamiltonian flow with $\mu_2$,
\be
\mu_2=\sqrt{A^2(\l_2)+B(\l_2)C(\l_2)}=
\sqrt{\alpha^2+\sum_{j=1}^n
\left(\frac{H_j}{\l_2-a_j}+\frac{s_j^2}{(\l_2-a_j)^2}\right)},
\ee
as the Hamiltonian function,
\beq
\dot L(u)=-\mbox{\rm i}\{\mu_2,L(u)\}.
\label{rlax1}
\eeq

This means that the two-point map discretizes a one-parameter
family of flows. Having chosen the parameter $\l_2$ to be equal
to any of the poles of the Lax matrix (parameters of the model)
 $a_j$, $j=1,\ldots,n$, the map leads to $n$ different maps,
each discretizing the flow with the corresponding Hamiltonian
$H_j$, $j=1,\ldots,n$. Indeed, take the limit $\l_2\rightarrow a_j$,
\beq
\l_2=a_j+\varepsilon,\qquad \varepsilon\rightarrow0.
\eeq
Then we have
\beq
\mu_2=\frac{s_j}\varepsilon+\frac{H_j}{2s_j}+O(\varepsilon),
\eeq
and in this limit the Lax equation (\ref{rlax})--(\ref{rlax1}) turns
into
\beq
\dot L(u)=-\frac{\mbox{\rm i}}{2s_j}\,\{H_j,L(u)\}
=\left[L(u),\frac{1}{2s_j(u-a_j)}\,\pmatrix{s_j^3&s_j^-\cr
s_j^+&-s_j^3}\right].
\eeq

Let us denote a collection of these maps by $\{B_{P_1}^{H_j}\}_{j=1}^n$.
The map $B_{P_1}^{H_k}$ discretizes the flow governed by the
Hamiltonian $H_k$ with $\l_1-a_k$ playing the role of the discrete
time-step parameter. The map (and its inverse) is defined by the
two-point matrix $M(u)$ (\ref{matrixM}) with the following expressions for the
variables $x$ and $X$:
\beq
x=\frac{A(\l_1)-\mu_1}{B(\l_1)}=\frac{\tilde{\tilde s}_k^{\;3}-s_k}
{\tilde{\tilde s}_k^{\;-}}\,,
\eeq
\beq
X=\frac{(a_k-\l_1)B(\l_1)s_k^-}{B(\l_1)(s_k^3+s_k)-s_k^-(A(\l_1)-\mu_1)}
=\frac{(a_k-\l_1)\tilde{\tilde B}(\l_1)
\tilde{\tilde s}_k^{\;-}}{\tilde{\tilde s}_k^{\;-}\left(\tilde{\tilde
A}(\l_1)+\mu_1\right)
-\tilde{\tilde B}(\l_1)\left(\tilde{\tilde s}_k^{\;3}-s_k\right)}\,.
\eeq

All these maps are explicit Poisson maps, preserving Hamiltonians and having
the spectrality property with respect to the pair of variables $(\l_1,\mu_1)$.

\newsection{Concluding remarks}
\setcounter{equation}{0}
\noindent
One of the very important branches of the theory of finite-dimensional
integrable systems is the area of discrete-time integrable systems.
The interest to this area was revived in the beginning of 90's by
Veselov in the series of works (see \cite{Ves}). He defined
integrable Lagrange correspondences as discrete-time analogs of
integrable continuous flows, clarified their geometric
meaning as finite shifts on Jacobians and gave several
important examples.
Since then the subject has got a boost and has been
developed further by many authors. It has not been our
intention to give a review of many important recent
contributions made to the area, because it would require
much more space. Instead, here we only mention in brief
the main features of a new recent approach to constructing
integrable maps which was introduced in
\cite{KS1,KS2,S}, developed in \cite{KV}, which
has also been used in this paper and which will be referred to
as B\"acklund transformations for finite-dimensional integrable
systems.

One of the new features of this approach to
disrete-time integrability
is {\it the spectrality property} which is a projection on the
classical case of the famous (quantum) Baxter equation.
It was discovered on the examples of Toda lattice and elliptic
Ruijsenaars system in \cite{KS1} and was generalized to the
integrable case of the DST model in \cite{KS2}. Later on, it was
observed that the property is universal and that, in effect,
it gives a canonical way of parameterising the corresponding
shift on the Jacobian which is characterized by adding a point
$(\l,\mu)$ to a divisor of points on the spectral curve $\Gamma$
(cf. \cite{KV}).

A direct consequence of the spectrality property is the {\it explicitness}
of the constructed maps. This new point, which is an obvious advantage
because explicit iterative maps are much more useful than implicit maps
(given as a system of non-linear equations), was clearly demonstrated
in \cite{KV}. This new aspect of constructing {\it explicitly given maps}
has been also adopted and illustrated in detail in the present paper.

There were several examples of explicit maps known before,
like McMillan's map, but all those cases were exceptional,
for in the generic situation, according to Veselov's approach,
integrable Lagrange/Poisson correspondences are
multi-valued maps, i.e. {\it correspondences} rather than maps.
Using the spectrality property as extra data allows
to overcome this drawback and
to construct discrete-time integrable flows as genuine maps.

Another new feature of the proposed construction of integrable
time-discretizations is an identification of the most elementary,
one-point, basic map and construction of composite maps,
like the two-point map, as compositions of the one-point map and
its inverse. The choice of the matrix $M(u)$ (\ref{M})
generating the
one-point map is dictated by the algebraic considerations
explained in \cite{S}. In brief, the matrix $M(u)$ should
be a simple $L$-operator of the quadratic algebra,
\beq
\{L_1(u),L_2(v)\}=[r(u-v),L_1(u)L_2(v)],
\label{rrr}
\eeq
with the same rational $r$-matrix (\ref{rr}) as in the linear algebra
(\ref{r}). The number of zeros of the $\det M(u)$ is the number
of essential B\"acklund parameters, so that the matrix $M(u)$ in (\ref{M})
is one-point and the matrix $M(u)$ in (\ref{matrixM}) is two-point.
The fact that the right ansatz for the matrix $M(u)$ obeys the algebra
(\ref{rrr}) usually garanties that the resulted map will be Poisson,
see \cite{S} for details.

In the present paper we have observed a new `spectrality'
property of the basic one-point map with respect to the
parameter $\gamma$ in
\beq
\det M(u)=\gamma(u-\l).
\eeq
We have also shown that the two-point map factorises
to two one-point maps.

The two-point map constructed above is probably most general
map for the considered sl(2) Gaudin model, meaning
that it gives a discretization
of continuous flows given by any Hamiltonian $H_j$, $j=1,\ldots,n$,
from the spectral curve,
\beq
v^2=A^2(u)+B(u)C(u)=\alpha^2+\sum_{j=1}^n
\left(\frac{H_j}{u-a_j}+\frac{s_j^2}{(u-a_j)^2}\right).
\eeq
So, at least in principle, any other integrable map for this
model should be a function of the $n$ maps constructed
in this paper.

There is no established name for integrable maps
with all the qualities mentioned above, namely: i) spectrality,
ii) explicitness, iii) Poissonicity, iv) limits
to continuous flows,
v) preservation of the same integrals as for the continuous flows
which these maps discretize. We are using for them the same name,
{\it B\"acklund transformations}, as
was used in the references \cite{KS1,KS2,KV,S}.

The application of the constructed maps as exact
numerical integrators of the continuous flows
is considered in \cite{me}.

\section*{Acknowledgements}

VBK wishes to acknowledge the support from the EPSRC Advanced
Research Fellowship AF/100072. The hospitality and support
of the Universita degli Studi `Roma Tre' extended to VBK during his
visit to Rome in 1998 where part of this work was done is kindly
acknowledged too.


\end{document}